**Electron density modulation in monolayer MoS$_2$ along the phase transition of a relaxor ferroelectric substrate**

David Hernández-Pinilla[1,2], Dennis Cachago[1], Yi An Xia[1], Guillermo López-Polín[1,3], Mariola O Ramírez[1,2,3], Luisa E. Bausá[1,2,3]

[1]Dept. Física de Materiales, Universidad Autónoma de Madrid, Spain
[2]Instituto de Materiales Nicolás Cabrera (INC), Universidad Autónoma de Madrid, Spain
[3]Condensed Matter Physics Center (IFIMAC), Universidad Autónoma de Madrid, Spain

**ABSTRACT**

The integration of transition metal dichalcogenides (TMDs) with ferroelectric substrates is a powerful strategy to modulate their electronic and optical properties. However, the use of relaxor ferroelectrics for this purpose remains unexplored. Here, we demonstrate a reversible photoluminescence (PL) and charge density modulation of monolayer MoS$_2$ on a Sr$_{0,61}$Ba$_{0,39}$Nb$_2$O$_6$ (SBN) substrate, a prototypical relaxor ferroelectric. The smearing of the phase transition in SBN enables continuous tuning of MoS$_2$ electronic properties over a broad temperature range (30–90°C). A pronounced PL enhancement occurs as the substrate transitions from ferro-to-paraelectric phase due to the vanishing spontaneous polarization (P$_S$) and the consequent change in charge balance at the MoS$_2$/SBN interface. Moreover, thermal hysteresis in the electron density modulation is observed during heating and cooling cycles. These findings highlight the potential of relaxor ferroelectrics as reconfigurable platforms for electron doping and light-emission control in 2D materials, opening avenues for temperature-responsive optoelectronic and nanophotonic applications.

**INTRODUCTION**

Transition metal dichalcogenides (TMDs), such as $MoS_2$, have attracted significant attention due to their exceptional characteristics, including high carrier mobility, tunable band gaps, and light-matter interaction, which make them ideal for applications in photodetectors, transistors, and energy storage [1-4]. The ability to control their properties through external stimuli further enhances their versatility in emerging technologies [5-8].

Indeed, the modulation of electron doping in TMDs has become a key focus to develop switching systems with tunable characteristics. To that end, different methods, including chemical doping, electrostatic gating, strain engineering, and the use of external light fields, have been recently proposed [9-11]. Additionally, due to the high sensitivity of monolayer TMDs (1L-TMDs) to the surrounding environment, the use of ferroelectric substrates has emerged as an effective approach for controlling their properties [12,13].

In this regard, combining 1L-TMDs with ferroelectric materials has gained significant attention bypassing the need for intricate fabrication techniques that could hinder their integration into optoelectronic devices. A number of hybrid systems combining 2D materials and ferroelectrics have been studied for applications in memories [14], or a variety of optoelectronic devices [15]. A prominent example of the extent to which the properties of ferroelectric substrates can be exploited is the use of lithium niobate ($LiNbO_3$) as an excellent platform for tuning the electronic properties of 2D materials [16]. Previous studies on 2D/$LiNbO_3$ heterostructures have shown domain-dependent electrostatic doping [17] and explored the luminescent properties of 1L-TMDs, highlighting the ferroelectric control of excitonic quasiparticles [18,19]. A recent study identified a polarization-dependent, ferroelectrically induced photodoping process [20], and more recently, the effect of pyroelectricity on the electronic doping of 2D materials has been reported by optical measurements as well as by transport measurements [21,22].

In addition to $LiNbO_3$, a variety of other ferroelectrics have been used as gate substrates for 2D materials, including $BiFeO_3$ and PZT for dielectric memories [14,23,24] and $BaTiO_3$ as a ferroelectric dielectric for single-layer $MoS_2$ field-effect devices [25]. However, the integration of 2D materials with relaxor ferroelectrics has yet to be explored. In these systems, the Curie temperature ($T_C$) is not sharply defined. Instead, they exhibit a gradual transition from the ferroelectric to the paraelectric phase. This feature enables the study of various physical phenomena across the phase transition, as the material evolves from the ferroelectric to the paraelectric phase or vice versa [26-29], where pronounced changes of the relevant physical magnitudes are expected.

In this work, an $Sr_{0.61}Ba_{0.39}Nb_2O_6$ crystal (hereafter SBN) is used as an active substrate for a $MoS_2$ monolayer (1L-$MoS_2$). This crystal belongs to the attractive class of relaxor ferroelectrics [30] and, additionally, presents remarkable properties such as large pyroelectric and electro-optical coefficients, as well as favorable photorefractive and acousto-optic characteristics [31], potentially useful for integrated devices. Indeed, SBN has been proposed for a wide range of applications in photonics, data storage, or switching [32-35].

SBN has a tungsten-bronze structure and belongs to the *4mm* point group in its ferroelectric phase. In this phase, the $Nb^{5+}$, $Sr^{2+}$, and $Ba^{2+}$ cations are displaced with respect to the *m* symmetry plane, creating a series of dipoles aligned in the same direction along the c-axis. This alignment gives rise to spontaneous polarization in the material. Above $T_C$, SBN belongs to the *4/mmm* point group. In this phase, $Sr^{2+}$, $Ba^{2+}$, and 20 % of the $Nb^{5+}$ ions are positioned within the *m* symmetry plane, while the remaining 80 % of the $Nb^{5+}$ ions are symmetrically distributed on both sides of the plane. In this configuration, the cations no longer generate spontaneous polarization, and SBN becomes paraelectric [36,37].

The structure of SBN exhibits notable cationic disorder, with ion distribution occurring in a disordered manner across different coordination environments. This underlies the smearing of the phase transition, which occurs over a relatively broad temperature range [38]. In our work, the SBN crystal used as a substrate ($Sr_xBa_{1-x}Nb_2O_6$) corresponds to a composition of x = 0.61, which exhibits a relatively low $T_C$ around 70 °C [29].

More broadly, structural cation disorder and compositional fluctuations lead to local symmetry distortions, so that SBN crystals typically contain randomly embedded inclusions of small, spontaneously polarized regions. In fact, as previously reported, SBN exhibits a disordered distribution of anti-parallel, needle-like ferroelectric domains, with the longest dimension aligned along the optical c-axis. Some relevant applications of SBN crystals, such as beam fanning reversals, polarization-based adjustable memories and frequency conversion, are related to the presence of this domain distribution [39, 40].

In short, these peculiar characteristics, along with the low $T_C$ value, highlight the relevance of exploring SBN as a substrate for modulation and switching processes in 2D materials.

The objective in this work is to investigate the influence of the ferro-to-paraelectric transition on the optical and electronic properties of 1L-$MoS_2$ that has been transferred onto the polar surface of an SBN substrate. To this end, we study the evolution of the photoluminescence (PL) of 1L-$MoS_2$ around $T_C$ of SBN, within the 30–90°C range in which the relaxor phase transition takes place. As the substrate undergoes from the ferroelectric to the paraelectric phase, a significant enhancement in the 1L-$MoS_2$ emission intensity is observed. This enhanced emission is attributed to the vanishing of $P_s$ as the substrate transitions to the paraelectric phase, which alters the polarization/screening-charge balance at the 1L-$MoS_2$/SBN interface, thereby modulating the carrier density in the 1L-$MoS_2$.

Moreover, the work demonstrates the reversibility of the modulation of the PL and electron doping along the substrate phase transition while featuring a hysteric behavior. This highlights the reconfigurable ability of the substrate to induce switchable functionality through its interaction with the 2D material. The findings show a robust strategy for electrostatic carrier modulation in 1L-TMDs, demonstrating the potential for temperature-responsive optoelectronic devices that do not require external electric fields. The approach offers a straightforward, reversible method for controlling electron doping and light emission, paving the way for advanced nanophotonic devices and sensing technologies with tunable characteristics.

## Results and discussion

**Figure 1a** shows a schematic representation of the system used in this work. It consists of a 1L-MoS$_2$ transferred onto the polar surface of a SBN relaxor ferroelectric crystal. Details on the sample preparation are provided in the Methods section.

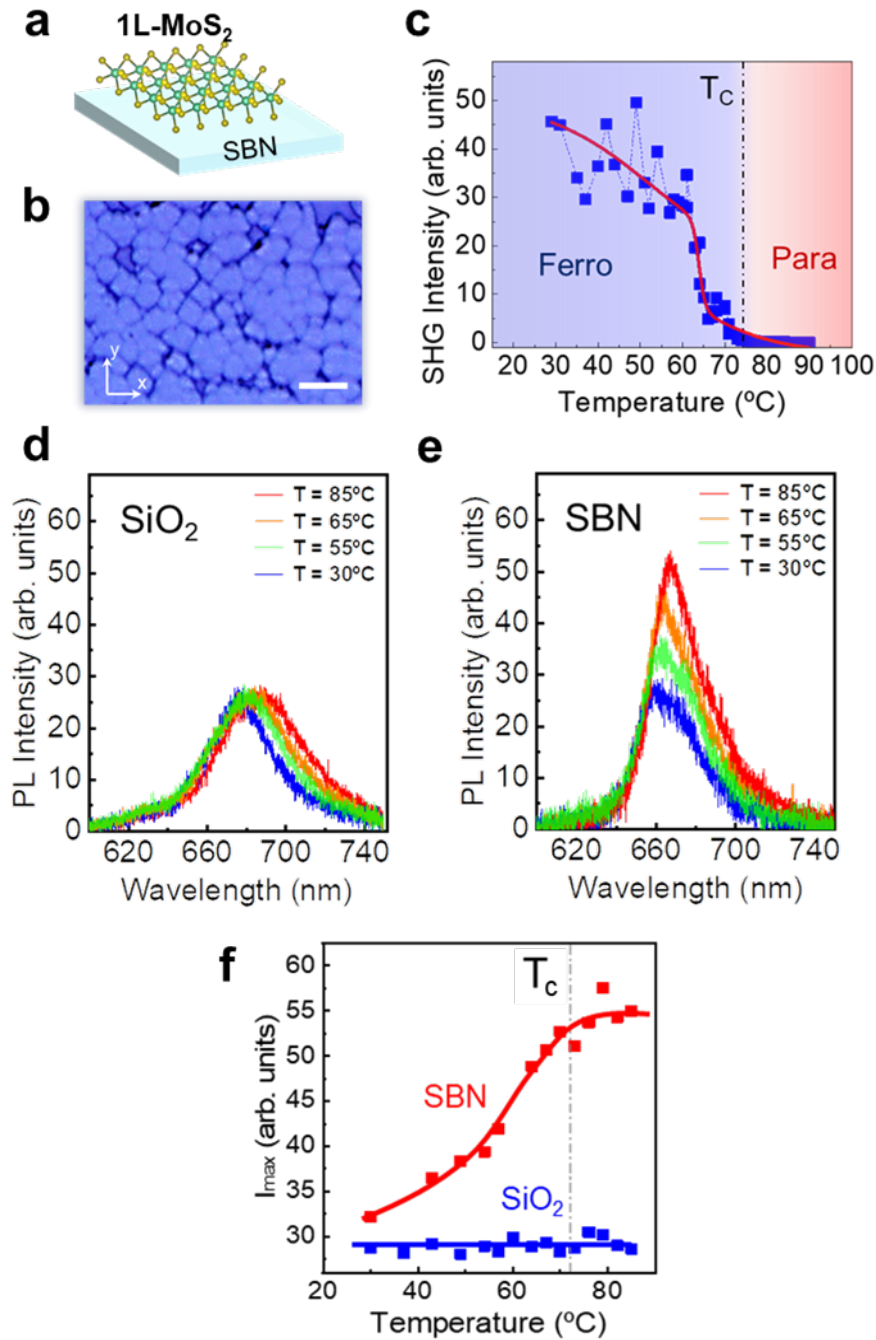

**Figure 1**. a) Schematics of the 1L-MoS$_2$ transferred on a SBN ferroelectric substrate. b) Optical image of the ferroelectric domain distribution on the a-b plane of the SBN crystal. Scale bar: 1µm. c) Evolution of the SHG intensity of SBN as a function of temperature around the relaxor ferro-to paraelectric phase transition. Evolution of the PL spectra of 1L-MoS$_2$ with increasing temperature on a d) SiO$_2$ and a e) SBN substrate. f) PL intensity evolution of 1L-MoS$_2$ on SiO$_2$ (blue) and SBN (red) as the temperature increases. The lines are guides to the eyes.

A prior characterization of the SBN substrate before MoS$_2$ deposition is of particular interest. On one hand, the characterization of the domain structures has been performed by optical microscopy. **Figure 1b** shows an optical image of the c- surface of the SBN crystal revealing a random distribution of ferroelectric domains in the a–b plane after selective chemical etching. The observed distribution is consistent with previous reports, which described needle-like ferroelectric domains with a quasi-square cross-section [26,41], with their longest dimension parallel to the optical c- axis. In our case, the average domain diameter is approximately 500 nm.

On the other hand, determining the phase transition temperature is also crucial in the context of this work. To that aim, non-invasive optical measurements have been used. In particular, second harmonic generation (SHG) has been employed, since investigating the evolution of quadratic optical nonlinear phenomena provides a reliable and direct means to track symmetry changes in the crystal. **Figure 1c** illustrates the evolution of the SHG signal of the SBN substrate as a function of temperature. At room temperature, the crystal is in its ferroelectric phase characterized by the noncentrosymmetric tetragonal space group P4bm, which allows quadratic nonlinear optical processes such as SHG. As the temperature increases, a noticeable change in the SHG intensity is observed, due to the transition towards the centrosymmetric paraelectric phase, where the quadratic nonlinear optical effects vanish. Namely, the SHG signal gradually decreases with increasing temperature in the 30–70 °C range, according to the relaxor character of SBN, until it reaches a value of nearly zero at around 73 °C, which we identify as the $T_C$ of our substrate. This value is similar to those previously reported for SBN crystals with the same stoichiometry (x = 0.61) [38,29].

Once the relaxor ferroelectric behavior has been established, we have analyzed the PL of 1L-MoS$_2$ on the SBN substrate as a function of temperature around the phase transition. To isolate the effects of the phase transition from those of temperature-induced changes in the radiative emission rate, the evolution of the PL of 1L-MoS$_2$ on SBN is compared to that on SiO$_2$, a system that does not undergo a phase transition. The temperature is varied from 30 °C to 90 °C, and PL spectra is measured at intervals of 5 – 10 °C. For clarity, only selected spectra at some specific temperatures are shown. **Figure 1d** and **1e** presents the emission spectra of both systems in the region of the A-exciton, measured under the same conditions as the temperature increases. The most significant difference between the two series of spectra lies in their intensity. While the PL of 1L-MoS$_2$/SiO$_2$ shows no significant changes in intensity with increasing temperature, the 1L-MoS$_2$/SBN system exhibits remarkable enhancement in emission, with the intensity at 90 °C reaching nearly twice that at room temperature. **Figure 1f** shows the comparison of PL intensity evolution of 1L-MoS$_2$ on SiO$_2$ and on SBN as the temperature increases. The difference in the PL intensity behavior of 1L-MoS$_2$ on SBN compared to SiO$_2$ is attributed to the SBN relaxor ferro-to paraelectric phase transition. The possibility of PL enhancement in MoS$_2$ using a substrate that undergoes an insulator-to-metal phase transition has been reported. However, the observed enhancement is attributed to constructive optical interference when the substrate becomes metallic [42].

A more detailed analysis of the spectra reveals further differences between both types of systems. In the PL spectra of **Figure 1d**, the emission from 1L-MoS$_2$/SiO$_2$ can be primarily attributed to trions (i.e., two electrons and one hole), with a band peaking at approximately 675 nm at room temperature [43], which redshifts as the temperature increases due to thermal-induced effects [44]. The predominantly trionic character of the emission is consistent with the intrinsic n-type doping of MoS$_2$ layers and the prominent electron doping provided by the SiO$_2$ substrate [45], which persists across the entire temperature range studied. In contrast, as observed in **Figure 1e,** the PL spectra of 1L-MoS$_2$/SBN reveal emission from two types of quasiparticles: excitons (i.e. electron-hole pairs) and trions, with peaks at around 658 nm and 675 nm, respectively, at room temperature. The thermal evolution of these spectra also exhibits a redshift. However, in contrast to the SiO$_2$ substrate, a significant intensity enhancement is observed at the highest temperatures. Moreover, as the temperature increases, a clear conversion of trion into neutral exciton emission occurs, as will be shown in further detail below.

The observed differences can be attributed to the substrate effect. Specifically, the different excitonic contribution in the room temperature spectrum of 1L-MoS$_2$/SBN with respect to that of 1L-MoS$_2$/SiO$_2$, can be linked to the effect of the ferroelectric substrate, which induce a modulation of the electron density in agreement with previous studies [20].

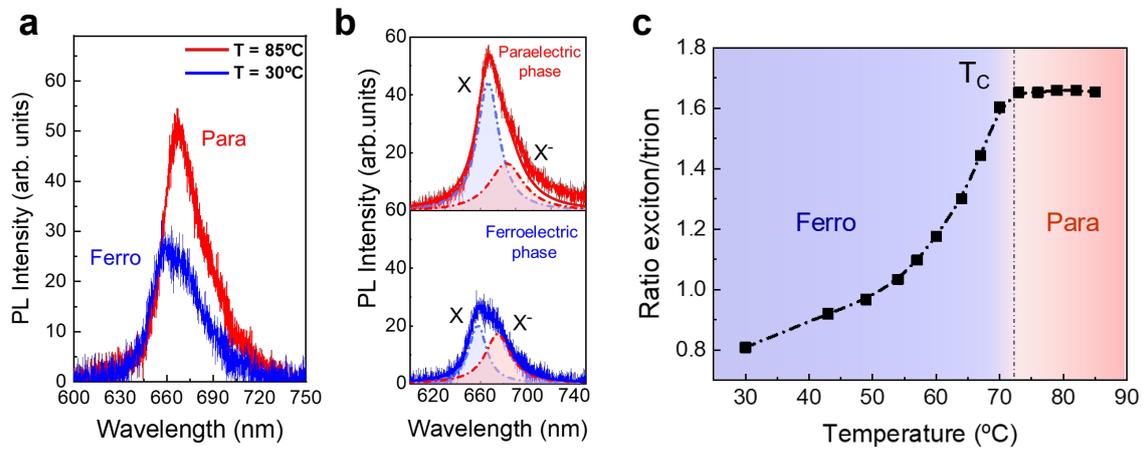

**Figure 2.** a) Comparison of the emission spectra of the 1L-MoS$_2$ monolayer in the ferroelectric and paraelectric phases of the SBN substrate. b) Deconvoluted PL spectra of 1L-MoS$_2$ showing the exciton and trion contributions in the ferroelectric and paraelectric phases of the SBN substrate. The labels X and X$^-$ refer to exciton and trion, respectively. c) Evolution of the exciton-to-trion intensity ratio of 1L-MoS$_2$ with increasing temperature across the ferroelectric to paraelectric transition of SBN. The line is a guide to the eyes.

To clearly visualize the modification of the 1L-MoS$_2$ PL occurring during the phase transition, we compare the spectra taken at both ends of the temperature range, in the ferro- and paraelectric phases, as shown in **Figure 2a.** There, the increase in PL intensity as the substrate transitions from the ferroelectric to the paraelectric phase is readily observed. These spectra have been deconvoluted into their exciton and trion contributions, using Lorentzian bands (**Figure 2b**). This analysis reveals that the increased emission intensity in the paraelectric phase is linked to the enhancement of excitonic emission, while the trion emission remains essentially constant. The evolution of the exciton-to-trion intensity ratio across the entire temperature range is shown in **Figure 2c**. As observed, this ratio increases monotonically with increasing temperature in the 30–70 °C range, indicating a gradual variation of the electron density of the monolayer up to the T$_C$, after which the exciton-to-trion intensity ratio remains practically constant. This result correlates well with the relaxor phase transition of the SBN substrate, which, according to Figure 1c, occurs around 73 °C.

The observed variations in 1L-MoS$_2$/SBN PL can be related to changes in the electronic density ($n_e$), which can be estimated from the spectral weight ratio of trions to excitons. Based on the mass action law, which describes the equilibrium between excitons, trions, and free electrons, $n_e$ can be determined from the PL spectra without requiring electrical measurements, as follows: [43]

$$\frac{I_{X^-}}{I_X} = \frac{n_e}{\eta_r C(T)} \qquad (1)$$

where $I_{X^-}$ and $I_X$ represent the intensity contribution of trions and excitons to the spectra, respectively. $\eta_r$ denotes the relative quantum yield of the exciton and the trion ($\eta_r = \eta_X/\eta_{X^-}$), while $C(T)$ is a function of temperature, which is given by

$$C(T) = \left(\frac{4 m_X m_e}{\pi \hbar^2 m_{X^-}}\right) \cdot k_B T \cdot e^{-E_b/k_B T} \qquad (2)$$

Here, $m_X$, $m_{X^-}$ and $m_e$ refer to the effective masses of excitons, trions and electrons, respectively, with values of $m_X = 0.8 m_0$, $m_{X^-} = 1.15 m_0$ and $m_e = 0.35 m_0$ for 1L-MoS$_2$, [22] where $m_0$ is the free electron mass. The trion binding energy is $E_b \simeq 20$ meV [46]. At 300 K, $C(T)$ takes a value of $4.86 \cdot 10^{12}$ cm$^{-2}$. The radiative emission rate of the exciton is significantly higher than that of the trion, and $\eta_r$ has been approximated to 20/3 [43]. This value has been kept fixed throughout the analysed temperature range.

**Figure 3a** shows the electron density (in µC/cm$^2$) in 1L-MoS$_2$ as a function of increasing temperature within the phase transition range of the SBN substrate. Around room temperature, the electron density reaches a value of around 14 µC/cm$^2$ (9 x10$^{13}$ e/cm$^2$), which gradually decreases up to T$_C$. For higher temperatures, the charge density remains practically constant in the studied range, at a value of approximately 8 µC/cm$^2$. The observed electron density evolution in 1L-MoS$_2$ can be associated with the gradual disappearance of P$_S$ in the SBN substrate as it undergoes the relaxor ferro-to paraelectric phase transition. As a consequence, throughout the phase transition, a reduction in the polarization charge of the substrate occurs, modifying the charge balance at the 1L-MoS$_2$/SBN interface, and thus, the electron doping of the monolayer. In contrast, in the paraelectric phase, P$_S$ is zero, and the surface charge of the substrate is no longer altered, thereby maintaining the charge at the 1L-MoS$_2$/SBN interface mostly constant.

However, it is important to consider that the SBN substrate exhibits a ferroelectric domain distribution that results in regions with oppositely oriented polarization. In this scenario, as temperature increases, 1L-MoS$_2$ is expected to undergo the same charge density change (Δn$_e$) on SBN regions with opposite polarization orientations, but with the corresponding opposite sign.

**Figure 3b** sketches the variation in the interfacial screening and polarization charge associated with the disappearance of P$_S$ on both domain orientations (P$_{up}$ and P$_{down}$). The net electron doping of 1L-MoS$_2$ is not represented. As the temperature increases, the decrease in P$_S$ reduces the positive or negative polarization charge of the substrate depending on the domain orientation, leading to a charge imbalance at the 1L-MoS$_2$/SBN interface. As a result, the electron doping is modulated, increasing the electron density on the positive domains and decreasing it on the negative ones.

A point that should be addressed concerns the poli-domain structure of the SBN substrate. The obtained PL contain the contribution of 1L-MoS$_2$ from both type of domains with oppositely oriented polarizations (P$_{up}$ and P$_{down}$). In fact, the recorded spectra are an average of the expected emissions from 1L-MoS$_2$ on each type of domain orientation. This averaging effect arises from the domain size in SBN (average of 500 nm) and the spot size of our experiments (~2 μm), which prevents the detection of contributions of MoS$_2$ emission on isolated single domains. In addition, it should be noted that, while exciton PL intensity is highly sensitive to variations in electron doping, the PL intensity of trions exhibits an almost negligible dependence on electron doping [47]. Therefore, although 1L-MoS$_2$ on P$_{up}$ domains experiences an increase in negative doping, the expected increase in PL trion from P$_{up}$ is minimal. Meanwhile, the observed PL exciton enhancement, which should originate on the P$_{down}$ domains, is also affected by the decreasing contribution from the P$_{up}$ domain. Since both domains contribute simultaneously to the PL, the resulting signal is averaged, masking the intrinsic $\frac{I_{X-}}{I_X}$ ratio on each domain and, consequently, the full extent of their respective doping variations. As a result, the net charge variation in 1L-MoS$_2$ extracted from our PL measurements is smaller than changes reported for P$_S$ in close SBN compositions from electrical measurements in single-domain crystals (which range from 15 to 20 μC/cm$^2$ [48]).

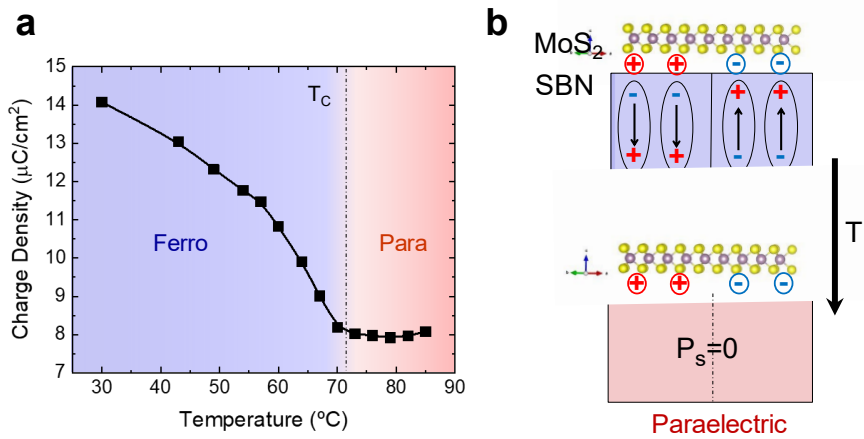

**Figure 3.** a) Evolution of electron doping in 1L-MoS$_2$ as temperature increases across the phase transition of the SBN substrate. b) Schematics of the cross section of the 1L-MoS$_2$/SBN interface. Modification of the screening/polarization charge balance at the 1L-MoS$_2$ interface as the temperature increases. The net electron doping of 1L-MoS$_2$ is not represented. Top: system in the ferroelectric phase with arrows representing spontaneous polarization. Bottom: system in the paralectric phase after heating. Screening charges are represented by circles.

Similar Δσ values have been obtained for a different 1L-MoS$_2$ with higher initial electron doping. This highlights that the electron doping modulation is independent of the starting doping, being determined by the variation in P$_S$. As a figure of merit, the evolution of the magnitude characterizing the electron modulation, Δσ, is shown as a function of temperature for a full thermal cycle: heating up to T$_C$ followed by cooling down to room temperature (see **Figure 4**). Upon increasing the temperature, the gradual decrease in Δσ is observed until it becomes zero at approximately 73 °C. On the other hand, the values corresponding to the temperature decrease show that the system returns to its initial state upon reaching room temperature. This result demonstrates the possibility of a reversible modulation of the optical and electronic properties of the 2D material, highlighting the substrate ability to act as a reconfigurable platform.

In addition, the most notable observation is the presence of thermal hysteresis in the modulation of electron density. That is, the transition temperature upon heating differs from that upon cooling, leading to a shift in the observed transition temperature. This hysteretic behavior is inherent to SBN, as demonstrated in a variety of properties analyzed in this crystal along the phase transition [28, 29, 49] and has been related to defects and cationic disorder in the lattice that introduces local compositional disorder as well as to domain reconfiguration and pinning. In our case, the hysteresis of SBN confers bistability to the electronic and optical properties of the 2D material.

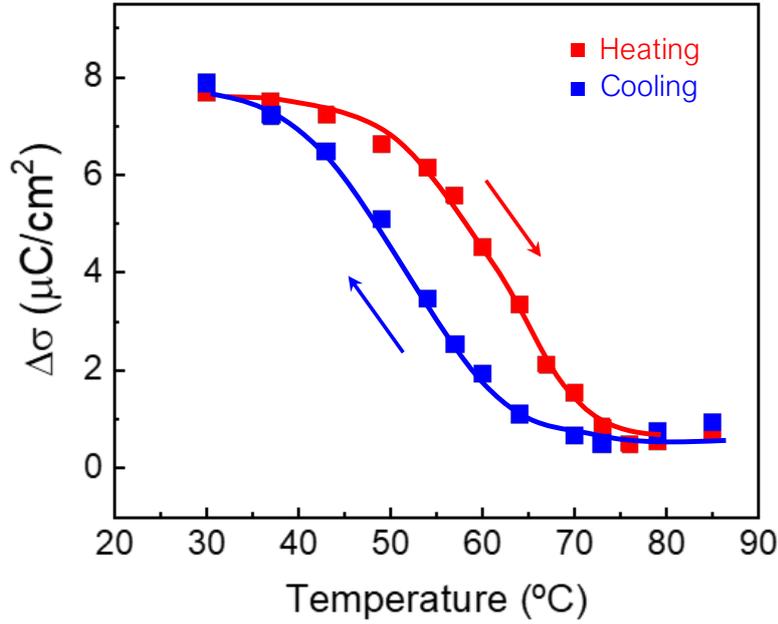

**Figure 4.** Temperature dependence of the net change in electron density of 1L-MoS$_2$ exhibiting hysteresis between heating (red squares) and cooling (blue squares) processes around the phase transition of the SBN substrate. Solid lines are a guide to the eye.

In order to analyze whether other spectroscopic parameters of the 1L-MoS$_2$ PL are affected by the phase transition, we have studied the evolution of the spectral position and full width at half maximum (FWHM) of the excitonic band as a function of temperature for heating and cooling.

**Figure 5a** shows the evolution of the exciton energy in the 30–90 °C range upon heating and cooling. As observed, no changes or singularities that can be associated with the phase transition of the substrate are detected, nor is there any evidence of a hysteresis cycle. Moreover, the energy of the exciton decreases with increasing temperature, following the expected redshift that can be well described by a standard hyperbolic cotangent relation [50]

$$E_G(T) = E_G(0) - S\langle\hbar w\rangle \left\{ \coth\left[\frac{\langle\hbar w\rangle}{2k_B T}\right] - 1 \right\} \quad (3)$$

which characterizes the temperature dependence of the optical band gap energy, $E_G(T)$, with respect to its value at 0 K, $E_G(0)$. $S$ is the electron-phonon coupling constant, and $\langle \hbar w \rangle$ is the average energy of the relevant phonons, taken as $24\ meV$ [50]. The solid line in Figure 5a corresponds to the fit of the experimental data to equation (3) with $E_G(0) = 1.97 \pm 0.01\ eV$ and $S = 2.94 \pm 0.20$, consistent with previously reported values for MoS$_2$ [43, 51].

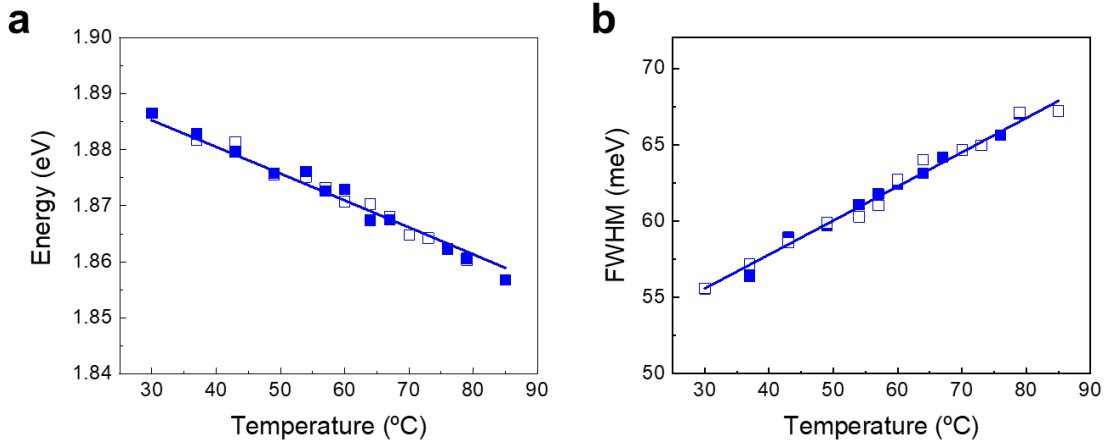

**Figure 5.** a) Spectral positions of the exciton in 1L-MoS$_2$/SBN during heating (full squares) and cooling (open squares). The data are fitted to the semi-empirical O'Donnell-Chen equation (ref. 50). b) FWHM of the exciton band upon heating (full squares) and cooling (open squares). The solid line represents a fit to equation 4.

The evolution of the FWHM is displayed in **Figure 5b**. Again, no abrupt changes or hysteresis are detected in this parameter as the SBN substrate undergoes the ferro-to paraelectric phase transition and the evolution of the bandwidth with increasing temperature follows a phonon-induced broadening that can be approximated by the phenomenological expression [51]

$$\gamma = \gamma_0 + c_1 T + \frac{c_2}{e^{\left(\frac{\langle \hbar w \rangle}{k_b T}\right)} - 1} \qquad (4)$$

where $\gamma_0$ and $c_1$ corresponds to the linear increase due to acoustic phonons, $c_2$ describes the strength of the phonon coupling and $\langle \hbar w \rangle$ is the average phonon energy. The fitting values of $\gamma_0$ and $\langle \hbar w \rangle$ have been taken as $\gamma_0 = 4\ meV$ and $\langle \hbar w \rangle = 24\ meV$ [51]. The curve in Figure 5b corresponds to the fit of the experimental data to equation (4) with $c_1 = 0.073 \pm 0.015\ meV$ and $c_2 = 45 \pm 7\ meV$, in agreement with previous works [51].

The energy positions and linewidths that characterize the emitting species follow the expected behavior for 1L-MoS$_2$ as a function of temperature, with no discontinuities or thermal hysteresis. This indicates that the 2D material itself is not influenced by the phase transition beyond the changes in electronic doping, which are primarily reflected in variations in the relative emission intensities of the quasiparticles.

**Conclusions**

In this work, we have demonstrated the reversible modulation of the PL and charge density of 1L-MoS$_2$ by utilizing the ferro-to-paraelectric phase transition of a relaxor ferroelectric SBN substrate coupled to the 2D material. This transition enables continuous and tunable control over the electronic properties of MoS$_2$ within a broad temperature range (30–90 °C). The gradual nature of the phase transition in SBN provides a unique platform for smooth, electrostatic doping of the 2D material, ensuring an adjustable charge carrier density. Furthermore, the inherent thermal hysteresis of the relaxor ferroelectric substrate induces a hysteretic behavior in both the PL and doping concentration of 1L-MoS$_2$, presenting exciting opportunities for applications in memory storage, artificial synaptic devices, and sensors that require reversible and temperature-responsive behavior. This ability to modulate properties via an intrinsic phase transition, without the need for external electric fields or mechanical stress, positions this approach as a promising strategy for integrating 2D materials with ferroelectric substrates in future optoelectronic and nanophotonic applications.

Finally, it is important to highlight that this work demonstrates the potential of relaxor ferroelectrics as a novel class of materials for reconfigurable platforms, offering a straightforward and effective method to gradually control the electronic properties of 2D materials. The findings open avenues for designing next-generation devices by a simple, robust approach for tuning light-emission and carrier density in monolayer TMDs, contributing to the ongoing development of advanced optoelectronic technologies. Moreover, the whole analysis has been carried out by using optical probes, emphasizing the adopted approach as a simple and effective way to modulate the electronic properties of 2D materials without relying on chemical, electrical or mechanical modifications.

**Methods**

**Sample Preparation**

A single congruent SBN *x*=0.61 crystal was grown by the Czochralski technique. The crystal contained a small amount of Yb$^{3+}$ (0.6 at% referred to Nb$^{5+}$ atoms). Plate samples were cut with the c-axis oriented perpendicular to the main face and then polished to achieve optical quality. Optical microscopy images were obtained in reflection mode using an Olympus BX51 microscope equipped with an Olympus DP72 camera. MoS$_2$ flakes were mechanically exfoliated from a bulk crystal, deposited onto a polydimethylsiloxane sheet, and subsequently deterministically transferred [52] at room temperature onto the main face of the SBN substrate. The monolayer character of the transferred MoS$_2$ flake was confirmed by differential micro-reflectance spectroscopy. An SiO$_2$ glass substrate was also used as a reference for the thermal evolution of the PL in 1L-MoS$_2$.

**Optical Measurements**

PL measurements were performed using a custom laser scanning confocal microscope (Olympus BX41), equipped with a motorized XY stage. An Ar$^+$ laser was used as excitation light source. Excitation power was fixed at 13.2 kW/cm$^2$ to avoid photodoping effects [20] and over-heating the sample. For SHG, the same confocal microscope system was employed. A femtosecond-pulsed Ti:sapphire laser (Spectra Physics Model 177-Series), operating at a wavelength of 800 nm, was used as the fundamental excitation source for SHG. In both types of optical measurements, the incident beam was focused on the sample using a 50X objective lens (NA = 0.35). The emission signals were collected in a backscattering geometry through the same objective by an optical fiber coupled to a Horiba iHR 550 monochromator. A Peltier-cooled Horiba Synapse CCD was used for the detection. To induce phase transitions during the PL experiments, the 1L-$MoS_2$/SBN sample was mounted on a heating platform (Linkam PE120) controlled by a T96 Peltier linkPad. Temperature was measured by means of chromel-alumel thermocouple place on the lateral face of the sample. Each emission spectrum was recorded once a steady-state temperature (equilibrium) was reached.

## Author contributions

D.H.P: Methodology, formal analysis, funding acquisition and writing review & editing; D.C.: investigation, formal analysis, and writing review & editing; Y.A.X.: investigation, formal analysis; G.L.P.: investigation and preparation of samples; M.O.R.: methodology, visualization, funding acquisition, and writing review & editing. L.E.B: conceptualization, supervision, funding acquisition, writing– original draft, and writing– review & editing.  All co-authors commented and provided inputs on the manuscript.

## Data availability
The data that support the findings of this study are available from the corresponding author upon reasonable request.

## Conflicts of interest
 The authors declare no conflict of interest.

## Acknowledgements

D.H.P, D.C, M.O.R, and L.E.B. acknowledge funding from the Spanish State Research Agency MICIU/AEI/10.13039/501100011033 under grant PID2022-137444NB-I00 and from Comunidad de Madrid and Universidad Autónoma de Madrid under grant SI4/PJI/2024-00217. G.L.P. acknowledges financial support from the Spanish State Research Agency under grant PID2022-138908NB-C32 and the the Ramón y Cajal contract RYC2023-044003-I. The authors acknowledge AEI under grant "Maria de Maeztu" Programme for Units of Excellence in R&D CEX2023-001316-M.